\documentclass{IEEEtran}
\usepackage[normalem]{ulem}
\usepackage{graphicx}
\usepackage{xcolor}
\usepackage{amssymb}
\usepackage{multirow}
\usepackage{multicol}
\usepackage{diagbox}
\usepackage{tabularx}
\usepackage[cmex10]{amsmath}
\usepackage{url}
\usepackage{pifont}
\usepackage{dsfont}
\usepackage{amsmath}
\usepackage{cite}
\usepackage{subfigure}
\usepackage[absolute,showboxes]{textpos}

\TPMargin{3pt}

\newcommand{\copyrightstatement}{
    \begin{textblock}{0.84}(0.08,0.953) 
         \noindent
         \footnotesize
         \copyright 2023 IEEE. Personal use of this material is permitted. Permission from IEEE must be obtained for all other uses, in any current or future media, including reprinting/republishing this material for advertising or promotional purposes, creating new collective works, for resale or redistribution to servers or lists, or reuse of any copyrighted component of this work in other works. DOI: 10.1109/TVT.2023.3309034
    \end{textblock}
}
\begin{document}
\copyrightstatement
%
\title{Cooperative Orbital Angular Momentum Wireless Communications}
%
%
%
\author{Ruirui~Chen, Wenchi~Cheng,~\IEEEmembership{Senior Member,~IEEE}, Jinyang~Lin, and Liping~Liang
\thanks{
This work is supported in part by Natural Science Foundation of Jiangsu Province under Grant BK20200650, in part by National Natural Science Foundation of China under Grant 62071472, in part by China Postdoctoral Science Foundation under Grant 2019M660133, in part by Program for ``Industrial IoT and Emergency Collaboration'' Innovative Research Team in CUMT under Grant 2020ZY002, in part by Fundamental Research Funds for the Central Universities under Grant 2019QNB01 and 2020ZDPY0304. (Corresponding author: Ruirui~Chen.)
\par Ruirui Chen and Jinyang Lin are with School of Information and Control Engineering, China University of Mining and Technology, Xuzhou, 221116, China (emails: rrchen, linjinyang@cumt.edu.cn). Wenchi Cheng and Liping Liang are with State Key Laboratory of Integrated Services Networks, Xidian University, Xi'an, 710071, China (e-mail: wccheng, liangliping@xidian.edu.cn).
}
}

\markboth{IEEE Transactions on Vehicular Technology, VOL. 73, NO. 1, JANUARY 2024}%
{Shell \MakeLowercase{\textit{et al.}}: Bare Demo of IEEEtran.cls for IEEE Journals}

\maketitle

\begin{abstract}
Orbital angular momentum (OAM) mode multiplexing has the potential to achieve high spectrum-efficiency communications at the same time and frequency by using orthogonal mode resource.
However, the vortex wave hollow divergence characteristic results in the requirement of the large-scale receive antenna, which makes users hardly receive the OAM signal by size-limited equipment. To promote the OAM application in the next 6G communications, this paper proposes the cooperative OAM wireless (COW) communication scheme, which can select the cooperative users (CUs) to form the aligned antennas by size-limited user equipment. First, we derive the feasible radial radius and selective waist radius to choose the CUs in the same circle with the origin at the base station. Then, based on the locations of CUs, the waist radius is adjusted to form the receive antennas and ensure the maximum intensity for the CUs. Finally, the cooperative formation probability is derived in the closed-form solution, which can depict the feasibility of the proposed COW communication scheme.
Furthermore, OAM beam steering is used to expand the feasible CU region, thus achieving higher cooperative formation probability.
Simulation results demonstrate that the derived cooperative formation probability in mathematical analysis is very close to the statistical probability of cooperative formation, and the proposed COW communication scheme can obtain higher spectrum efficiency than the traditional scheme due to the effective reception of the OAM signal.
\end{abstract}

\begin{IEEEkeywords}
Orbital angular momentum, spectrum efficiency, user, signal reception, cooperative formation probability.
\end{IEEEkeywords}

\IEEEpeerreviewmaketitle

\section{Introduction}
\IEEEPARstart{P}{lane} electromagnetic wave based wireless communications have almost made full use of traditional orthogonal resources (such as time, frequency, etc) \cite{1}. Virtual reality, high-definition video, etc, which are emerging applications, bring about the rapidly increasing wireless traffic demand for the next 6G communications \cite{2}. To further improve the spectrum efficiency, orbital angular momentum (OAM) is expected to achieve a breakthrough in the information transmission approach \cite{3}. OAM describes the characteristic of electromagnetic wave rotating around the propagation axis, which produces the phase wavefront of electromagnetic wave in a vortex shape \cite{4}. The electromagnetic wave with OAM is called vortex wave, and different OAM modes are orthogonal with each other \cite{5}. Therefore, as a new orthogonal resource, OAM mode can be utilized to enhance the access capability and improve the reliability of anti-jamming by using mode division multiple access and the mode hopping technique, respectively \cite{7,7a}. Particularly, with orthogonality, different OAM modes can be used to transmit multiple data streams simultaneously for the significant increase of spectrum efficiency in the next 6G communications \cite{6,8}.

 Since the light beam with OAM was discovered in 1992, it attracts great interest in wireless communications due to its high spectrum efficiency \cite{10}.
The authors in \cite{10A} proposed the adaptive optic technique to mitigate the turbulence effect for OAM based underwater optical communications.
To improve the effective coverage, the authors of \cite{10A1} investigated the beam and probabilistic shaping scheme for OAM based indoor optical communications.
Besides OAM based optical communications, OAM mode has been applied to wireless communications in the radio frequency (RF) domain \cite{10A2}.
The optimal uniform circular array (UCA) design was proposed to select the OAM mode and radius of receive UCA for OAM wireless backhaul communications \cite{10B}.
To improve the spectrum efficiency, the authors of \cite{10C} proposed an OAM mode allocation method, which provided adjacent cells with different carriers of different OAM modes.
To date, the research on OAM mode multiplexing has made great progress in wireless communications \cite{10a,10b}.
It is worth noting that in 2013, 4 Gbps successful transmission of uncompressed video over the 60 GHz OAM wireless channel was demonstrated in \cite{11a}.
In \cite{11}, two wireless signals were successfully multiplexed at the same frequency in OAM mode multiplexing based wireless communications for the first time.
The authors of \cite{12} analyzed the microwave based OAM mode multiplexing communications, which demonstrated that the high spectrum efficiency can be achieved by using the low complexity receiver for future wireless communications.
In OAM multiplexing communications, a coaxial separation and convergence scheme for multiple OAM modes was proposed, and the crosstalk between multiple OAM channels can be effectively reduced \cite{12a}.
To achieve high spectrum efficiency in sparse multipath environment, the authors of \cite{13} proposed a scheme that combines OAM mode multiplexing with orthogonal frequency division multiplexing (OFDM).
For further increase of the multiplexing gain, concentric UCAs were utilized to achieve the 100 Gbps data rate of OAM wireless communications \cite{14}. By using plane spiral OAM mode-group, a partial arc sampling receive scheme was proposed to realize OAM wireless communications, which can obtain robust performance in the multipath environment \cite{14a}. The authors in \cite{14b} proposed a shared aperture patch antenna, which had the simple circuit design and easy integration, to realize OAM wireless communications.

However, the transceiver antenna alignment, which requires the perfect alignment between the transmit and receive antenna arrays, imposes a crucial challenge for the high spectrum efficiency achievement of OAM wireless communications \cite{15,15a,15b,15c}. The authors in \cite{16} drew a conclusion that if the transceiver antenna is not aligned, the performance of OAM wireless communications deteriorates quickly.
For OAM wireless communications, the misalignment between the transmitter and receiver would cause interference among different OAM modes \cite{16a}.
To overcome this problem, the authors of \cite{17} studied the effect of misalignment on the channel crosstalk of OAM wireless communications, which can be relieved by selecting the proper aperture size and mode spacing of the transmitted OAM beams.
In \cite{17a}, the authors proposed the beam steering method, which used switches connected to array elements, to solve the aperture alignment of the transmitter and receiver in OAM wireless communications.
To avoid the performance deterioration caused by antenna misalignment, the authors of \cite{18} compared the analog beam steering scheme and digital beam steering scheme, and concluded that the baseband digital OAM transceiver with digital beam steering has a better performance for misaligned multi-mode OAM broadband wireless communications.
Considering the size and cost of UCA, the authors in \cite{19} developed an efficient technique for beam steering of arbitrary UCA sizes based on OAM modes. A hybrid mechanical and electronic beam steering approach was proposed in \cite{19.1} to maximize the channel capacity of OAM wireless communications.

Besides the transceiver antenna alignment, another challenge of the OAM application is that the size-limited user equipment is hard to receive the OAM signal at long distance transmission due to the hollow divergence.
Specifically, the hollow divergence of the vortex wave beam is the zero intensity circle around the beam propagation axis \cite{19A1,19A2}. The intensity distribution of the vortex wave beam is a ring, which is proportional to the OAM mode and increases as the transmission distance increases. In other words, the large OAM mode will result in serious hollow divergence, and the energy ring of the vortex wave beam will diverge with the increase of the transmission distance \cite{19B}. To reduce the divergence of the OAM beam, the authors of \cite{19.2} proposed the spherical Luneberg lens, which improved the transmission performance and reliability of OAM wireless communications.
Note that the great majority of research on OAM wireless communications is based on the centralized UCA \cite{19a,19b,19c,19d,19e}. The authors in \cite{20} showed that the spectrum efficiency of UCA-based OAM wireless communications is affected by the transmission distance and the size of UCA. However, the energy ring of the vortex wave beam will diverge as the transmission distance increases, thus making the receive UCA size big at long distance transmission \cite{21}. The hollow divergence leads to difficulty in receiving and utilizing the OAM signal.
Furthermore, the authors in \cite{22} proved that the hollow divergence of the vortex wave beam greatly increases as the carrier frequency decreases. Therefore, the easily realized communication scheme, which can efficiently receive the OAM signal, is urgently demanded for  size-limited user equipment to realize the OAM mode multiplexing, where the hollow divergence is serious in the RF domain.

Motivated by the OAM signal reception of size-limited user equipment, the cooperative OAM wireless (COW) communication scheme, which can select the cooperative users (CUs) to form aligned receive antennas, is proposed in this paper to overcome the vortex wave hollow divergence and achieve the high spectrum-efficiency OAM point-to-point transmission (especially for the long distance).
The main contributions of the paper are summarized as follows:

\begin{itemize}
  \item According to the intensity distribution of the vortex wave beam, the feasible radial radius and selective waist radius are derived to select the CUs in the same circle with the origin at the base station (BS). Then, we propose the COW communication scheme to adjust the waist radius to form the aligned receive antennas and ensure the maximum intensity for the CUs.
  \item To depict the feasibility of the proposed COW communication scheme, we derive the closed-form expression of cooperative formation probability, which is the probability that at least two CUs can be selected as one cooperative pair (CP).
  \item To expand the feasible CU region, OAM beam steering is used to compensate the phase distortion based on oblique angles, which can achieve higher cooperative formation probability.
  \item The proposed COW communication scheme achieves higher spectrum efficiency than the traditional scheme due to the effective signal reception by size-limited user equipment, and the derived cooperative formation probability in mathematical analysis is very close to the statistical probability of cooperative formation.
\end{itemize}

The rest of this paper is organized as follows. Section II presents the system model of OAM wireless communications. Based on the intensity distribution of the vortex wave beam, we derive the feasible radial radius and selective waist radius in Section III. In Section IV, we propose a COW communication scheme to realize the COW communications.
Section V presents the cooperative formation probability analysis, which can validate the feasibility of the proposed COW communication scheme.
The signal compensation and spectrum efficiency performance are analyzed for our proposed COW communication scheme in Section VI.
Section VII gives numerical simulation results to validate the theoretical analysis of the COW communication scheme.
The conclusion is drawn in Section VIII.

\section{System Model}
\begin{figure}[thp]
\setlength{\abovecaptionskip}{0.cm}
\setlength{\belowcaptionskip}{-0.cm}
\centering
\vspace{-0.25cm}
\setlength{\abovecaptionskip}{0.cm}
\setlength{\belowcaptionskip}{-0.cm}
\includegraphics[height=5.6in,width=9cm]{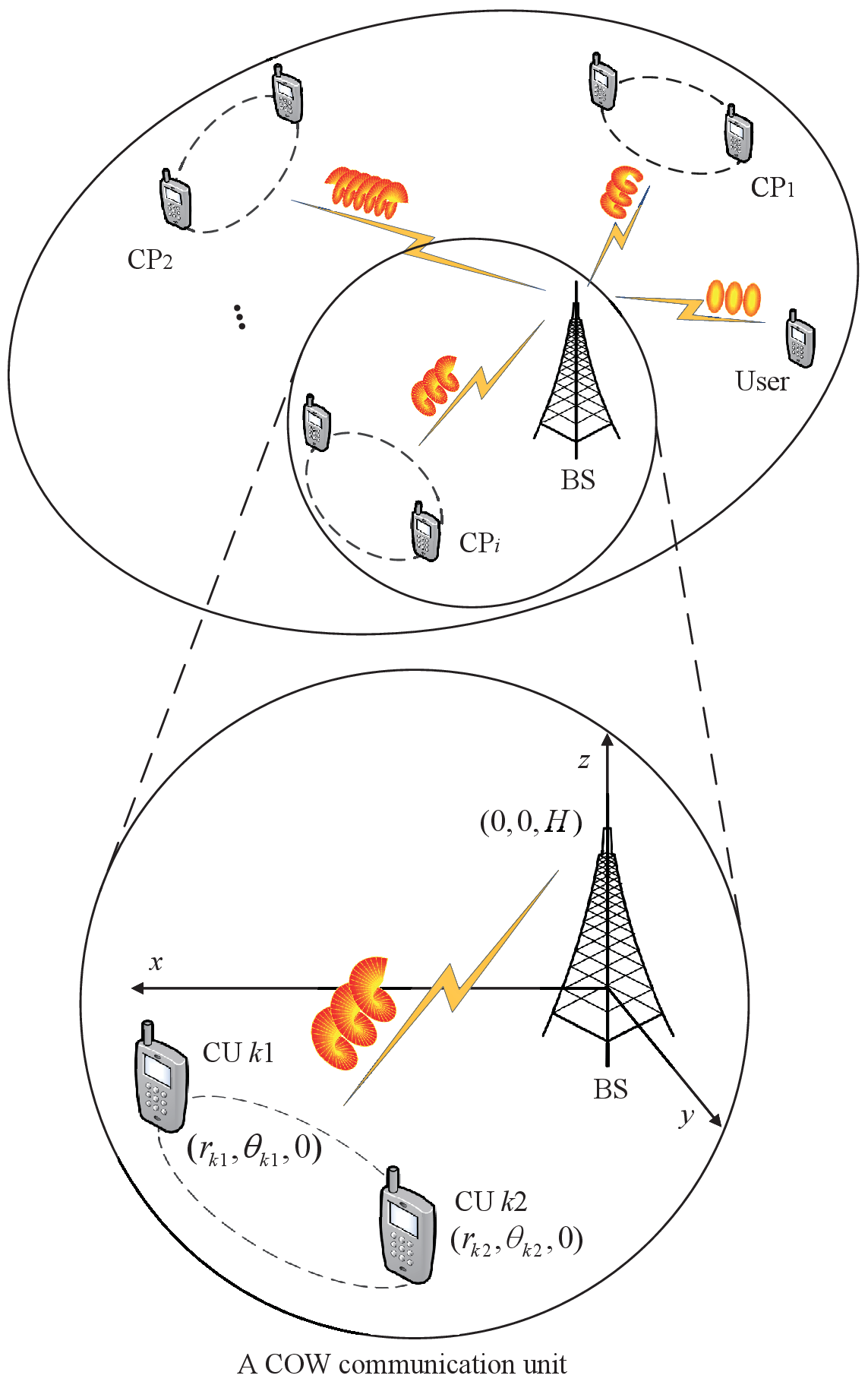}
\caption{\!The COW communications consist of the BS and multiple CPs. The zoomin figure at the bottom shows the detailed topology of a COW communication unit. The COW communications can be partitioned into a number of COW communication units. This implies that in the COW communication model, each user with size-limited equipment  uniquely belongs to one COW communication unit.}
\label{fig:aaaa}
\intextsep=1pt plus 3pt minus 1pt
\end{figure}

This paper assumes that there exists a BS and $K$ users in hotspot region, which is caused by the concert, festivals, and stadium games. The user set is represented by $\mathcal{K}=\{1,2\cdots,K\}$. The BS is equipped with the UCA, which has $M$ antenna elements denoted by the set $\mathcal{M}=\{1,2\cdots,M\}$. The users are all equipped with one antenna and randomly scattered on the ground.
The radius and central angle of UCA are $R$ and $\varphi=2\pi/M$, respectively. Furthermore, the user location information is known to the BS, which is easy for intelligent user equipment at present \cite{22a}.
As shown in Fig. \ref{fig:aaaa}, this paper considers the downlink COW communications, where there exists the Line-of-Sight (LoS) path and the channel experiences free-space path loss.

We define the two CUs as the $k_1$-th and $k_2$-th users. As shown in the bottom of Fig. \ref{fig:aaaa}, CU $k1$ and CU $k2$ are two users with size-limited equipment that can form a CP. A COW communication unit is defined as a unit that is composed of the BS and a CP. Thus, the COW communications can be divided into a number of COW communication units. This implies that in the COW communication model, each user uniquely belongs to a CP and also uniquely belongs to one COW communication unit. Based on the above analysis, we can focus on the COW communication unit to investigate the COW communications. Note that one user may carry multiple equipment that can cooperate with each other to form a CP. Furthermore, in hotspot region, there are many users clustered together, and incentive mechanism can be used to encourage user cooperation \cite{22b,22c}, which can guarantee the COW communications.

In the 3D cylindrical coordinate system, the BS with height $H$ is at the origin $(0, 0, 0)$, and the $z$-axis is perpendicular to the ground through BS. The UCA center of the BS and $k$-th $(k\in\mathcal{K})$ user are at the locations $(0, 0, H)$ and $(r_k, \theta_k, 0)$, respectively. The complex amplitude distribution of the Laguerre-Gaussian (LG) beam propagating along the $z$-axis is expressed as follows:
\begin{equation} \label{2e}
\begin{split}
U_{p,\ell }(r,\theta,z)=&\sqrt {\frac {2p!}{\pi w^{2}(z)(p+|\ell |)!}}  L_{p}^{|\ell |} \left ({\frac {2r^{2}}{w^{2}(z)}}\right) \left [{\frac {r\sqrt {2}}{w(z)}}\right]^{|\ell |}
\\&\exp (i\ell \theta)\exp \left [{\frac {i\beta r^{2}z}{2(z^{2}+z_{R}^{2})}}\right]\exp \left [{\frac {-r^{2}}{w^{2}(z)}}\right]
\\&\exp \left [{-i(2p+|\ell |+1)\tan ^{-1}\left ({\frac {z}{z_{R}}}\right)}\right],
\end{split}
\end{equation}
where $\theta$ is the azimuth angle, $\beta$ represents the wave number, $w_0$ is the waist radius at $z=0$, $w(z) = {w_0}\sqrt {1 + {{(z/{z_R})}^2}} $ denotes the radius of the beam at $z$, $z_R=\pi w_0^2/\lambda$ represents the Rayleigh range, $L_{p}^{|\ell |}(\cdot)$ denotes the associated Laguerre polynomial, and $(2p+|\ell |+1)\tan ^{-1}(z/z_R)$ is the Gouy phase. We can see that the LG beam is related with two parameters: the azimuthal index $\ell$ and the radial index $p$. The azimuthal index $\ell$ is the topological charge, which is also called OAM mode or vortex wave mode. The radial index $p$ is the number of radial nodes on the cross section of the LG beam intensity.

\section{Feasible Radial Radius and Selective Waist Radius}
The OAM beam is the LG beam with $p=0$. The intensity distribution is expressed as
\begin{equation} \label{2}
U_\ell(r,z)=\frac {2}{\pi w^{2}(z)|\ell |!} \left [{\frac {\sqrt {2}r}{w(z)}}\right]^{2|\ell |} \exp \left [{\frac {-2r^{2}}{w^{2}(z)}}\right].
\end{equation}
 Since $U_\ell(r,z)$ is a strictly convex function with respect to the radial radius $r$, we can obtain $r$ corresponding to the maximum $U_\ell(r,z)$ as follows \cite{19A1,19A2}:
\begin{equation} \label{3}
r_\textup{max}(z)=\sqrt{\frac{|\ell |}{2}}w(z).
\end{equation}
The maximum intensity is given as
\begin{equation} \label{6e}
U_\textup{max}(z)=\frac {2|\ell |^{|\ell |}\exp \left (-|\ell |\right)}{\pi w^{2}(z)|\ell |!}.
\end{equation}

According to (\ref{3}), we derive the waist radius $w_0$ corresponding to the optimal radial radius $r_\textup{max}(z)$ as
\begin{equation} \label{12e}
w_0=\sqrt{\frac{r_\textup{max}^2(z)}{|\ell |}\pm \sqrt{\frac{r_\textup{max}^4(z)}{|\ell |^2}-\frac{z^2\lambda^2}{\pi^2}}}.
\end{equation}
The following feasible radial radius should be satisfied to make the waist radius $w_0$ exist
\begin{equation} \label{11ee}
r_\textup{max}(z) \geq R_\textup{fea}(z) \triangleq \sqrt{\frac{z\lambda |\ell |}{\pi}}.
\end{equation}
Considering the size limitation of the BS, the smaller $w_0$ is the selective waist radius $w_0$, i.e.,
\begin{equation} \label{12ee}
w_0=\sqrt{\frac{r_\textup{max}^2(z)}{|\ell |}- \sqrt{\frac{r_\textup{max}^4(z)}{|\ell |^2}-\frac{z^2\lambda^2}{\pi^2}}}.
\end{equation}

To realize the $L$ OAM mode multiplexing communications, $L$ antennas uniformly locate at the $2\pi/a$ arc, and different OAM modes $l_n ~(n=1, 2, 3...N)$ satisfy $l_n-l_0=na$, where $l_0$ is a constant integer and $n$ is an arbitrary integer \cite{23}. Note that the user receives the OAM signal by using the size-limited equipment with only one antenna in general. We can see that $L$ CUs are required to achieve $L$ OAM mode multiplexing communications. Obviously, it is meaningless to realize the 1 OAM mode multiplexing communications. Furthermore, due to the random distribution of users in hotspot region, the more OAM modes are, the more difficult the realization of OAM multiplexing communications is. Therefore, we choose two CUs to achieve OAM wireless communications, which is the most practical case and can achieve 2 OAM mode multiplexing communications.

\section{ Cooperative OAM Wireless Communication Scheme}
In the 3D cylindrical coordinate system, the UCA center of the BS and $k$-th $(k\in\mathcal{K})$ user are at the locations $(0, 0, H)$ and $(r_k, \theta_k, 0)$, respectively.
The users are all uniformly scattered on the ground in hotspot region (i.e., horizontal plane $z=0$).
The location of the $m$-th $(m\in\mathcal{M})$ antenna element is at the location $(R, m\varphi, H)$.
The distance between the two CUs is denoted by $d_{{k_1},{k_2}}$.

Based on (\ref{3}), the two CUs should be close to the radial radius with maximum intensity. To this end, the two CUs should have the same distance to the BS and be located at the diameter of the circle with radius $r_\textup{max}(z^{\rm{c}})$, i.e., $r_{k_1} =r_{k_2} $ and $d_{{k_1},{k_2}}=2r_\textup{max}(z^{\rm{c}})$, where $z^{\rm{c}}$ denotes the distance from the UCA center of BS to the midpoint of the connection between the two CUs. Furthermore, the maximum communication distance is defined as $D_\textup{max}$ for the CUs.

Note that the propagation direction of $z^{\rm{c}}$ is the propagation axis of the LG beam, and the BS adjusts the waist radius $w_0$ to guarantee that the two CUs are located at the diameter of the circle with radius $r_\textup{max}(z^{\rm{c}})$. In other words, the waist radius $w_0$ is adjusted based on the radial radius $r_\textup{max}(z^{\rm{c}})=d_{k_1, k_2}/2$, and the transmission direction of the LG beam is from the UCA center of BS to the midpoint of the connection between the two CUs. Based on (\ref{11ee}), the condition $d_{{k_1},{k_2}}\!\geq\! 2R_\textup{fea}(z^c)$ should be satisfied to guarantee the existence of the waist radius.

The CU set and potential user (PU) set are defined as $\mathcal{R}_c$ and $\mathcal{R}_p$, respectively. The cardinalities of sets $\mathcal{R}_c$ and $\mathcal{R}_p$ are $|\mathcal{R}_c|$ and $|\mathcal{R}_p|$, which are respectively the numbers of users that belong to the CU and PU sets.
The ring with inner radius $r_{s}-\epsilon$ and outer radius $r_{s}+\epsilon$ is defined as the $r_{s}$ search ring, where $r_{s}$ denotes the search radius and $\epsilon$ is the extremely small value.
The search radius $r_{s}$ belongs to $(\epsilon,R_{\textup {BS}}-\epsilon)$, where $R_\textup{BS}$ is the maximum search radius (i.e., the BS coverage region).
Then, we propose the COW communication scheme, which can be summarized as follows.

\par \noindent\rule[0.25\baselineskip]{8.8cm}{0.5pt}
\uline{\textbf{Cooperative OAM wireless communication scheme}}
\par \noindent\emph{\textbf{Initialization}}:
\par \noindent~1) CU and PU sets $\mathcal{R}_c=\mathcal{R}_p=\emptyset$ and given user location
\par ~~\!$(r_k, \theta_k, 0)$;
\par \noindent~2) Set arbitrarily small value $\epsilon$ and search radius $r_{s} = \epsilon$;
\par \noindent\emph{\textbf{Iteration}}:
\par \noindent~3) \textbf{while} $(\mathcal{R}_c=\emptyset)$ \textbf{do}
\par \noindent~4) ~~\textbf{while} $(r_{s} \leqslant R_\textup{BS}-\epsilon)$ \textbf{do}
\par \noindent~5) ~~~~For $r_{s}$ search ring, find PU set $\mathcal{R}_p$;
\par \noindent~6) ~~~~\textbf{if}  $(\mid\mathcal{R}_p\mid \geq 2)$
\par \noindent~7) ~~~~~~~In PU set $\mathcal{R}_p$, select two CUs $k_1$ and $k_2$, which have
\par ~~~~~~~~\!the minimum $d_{{k_1},{k_2}} \in [ 2R_\textup{fea}(z^c), D_\textup{max}]$;
\par \noindent~8) ~~~~~~Add CUs $k_1$ and $k_2$ to the CU set $\mathcal{R}_c$;
\par \noindent~9) ~~~~\textbf{end if}
\par \noindent10) ~~~~Set $r_{s} = r_{s} + 2\epsilon$;
\par \noindent11) ~~\textbf{end while}
\par \noindent12) ~~Set $\epsilon=2\epsilon$;
\par \noindent13) \textbf{end while}
\par \noindent\rule[0.25\baselineskip]{8.8cm}{1pt}
\begin{figure}[thp]
\setlength{\abovecaptionskip}{0.cm}
\setlength{\belowcaptionskip}{-0.cm}
\centering
\vspace{-0.35cm}
\setlength{\abovecaptionskip}{0.cm}
\setlength{\belowcaptionskip}{-0.cm}
\includegraphics[height=2.2in,width=8.8cm]{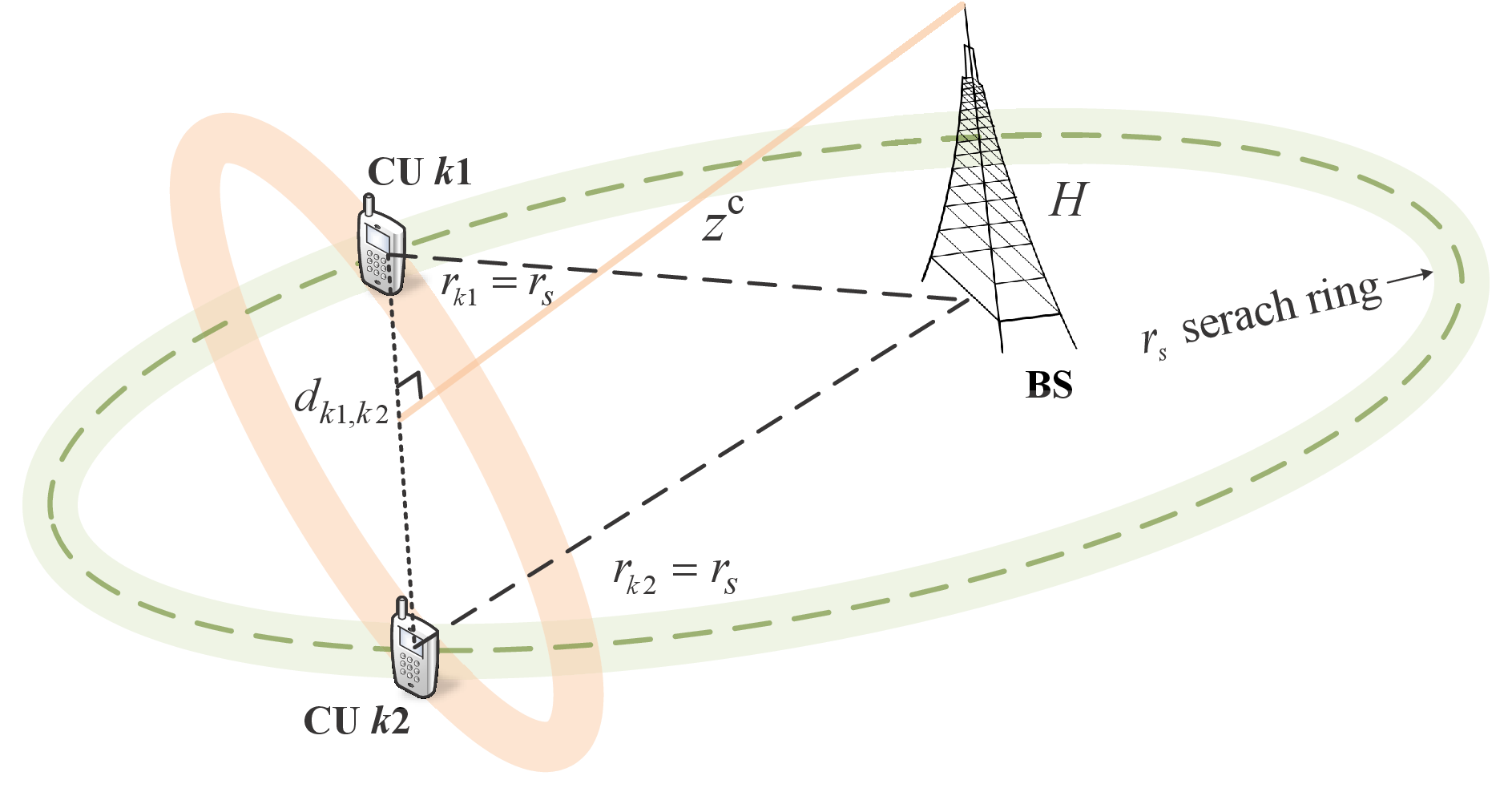}
\caption{\!The demonstration of proposed COW communication scheme}
\label{fig:aaaaa}
\intextsep=1pt plus 3pt minus 1pt
\end{figure}

Figure \ref{fig:aaaaa} gives the two selected CUs to illustrate the proposed COW communication scheme. The two CUs are within the $r_{s}$ search ring, which is the green circle with radius $r_{s}=r_{k_1}=r_{k_2}$. At the same time, the two CUs are located at the orange circle with diameter $d_{k_1, k_2}$ marked by the dotted line.
The distance from the UCA center of BS to the midpoint of the connection between the two CUs, which is denoted by $z^{\rm{c}}$, can be expressed as follows:
\begin{equation}\label{8abb}
{z^{\rm{c}}} = \sqrt {r_s^2 -\frac{ d_{k_1, k_2}^2}{4} + {H^2}} .
\end{equation}
The BS with height $H$ transmits OAM signals to the two CUs through the propagation direction of $z^{\rm{c}}$ (i.e., the propagation axis of the LG beam) marked by the orange solid line.

The PU set $\mathcal{R}_p$ can be found in the $r_{s}$ search ring according to the updated search radius $r_{s}$ (step 5).
If the number of the PUs is not smaller than two, the two CUs $k_1$ and $k_2$ that have the minimum $d_{{k_1},{k_2}} \in [ 2R_\textup{fea}(z^c), D_\textup{max}]$ are selected from the PU set $\mathcal{R}_p$ and added to the CU set $\mathcal{R}_c$ (step 4-11).
The above process should be repeated until the CUs are selected by the iteration of the $r_{s}$ search ring, i.e., execute the inner and outer while loop (steps 3-13).

For the complexity of proposed COW communication scheme, the maximum iteration numbers of the inner and outer loops are $\frac{R_\textup{BS}}{2\epsilon}$ and $\textup{log}_2(1+\frac{R_\textup{BS}}{\epsilon})$, respectively. The maximum iteration number of steps 5-10 is bounded by $\frac{R_\textup{BS}}{2\epsilon}\textup{log}_2(1+\frac{R_\textup{BS}}{\epsilon})$. For the sequential statements in steps 5-10, the complexity is determined by the highest complexity among the sequential statements, which can be expressed as a constant variable $W$. Therefore, the complexity of the proposed COW communication scheme is $O\left(\frac{R_\textup{BS}}{2\epsilon}\textup{log}_2(1+\frac{R_\textup{BS}}{\epsilon})W\right)$. Note that the complexity of the proposed COW communication scheme is proportional to $O\left(\frac{1}{\epsilon}\textup{log}_2(1+\frac{R_\textup{BS}}{\epsilon})\right)$. Based on the above analysis, the complexity of the proposed COW communication scheme can be bounded by $O\left(\frac{1}{\epsilon}\textup{log}_2(1+\frac{R_\textup{BS}}{\epsilon})\right)$, which is the tolerable complexity. Note that the BS has the ability to be equipped with multiple UCAs for covering all directions. Furthermore, the beam steering can be employed to expand the BS coverage and adjust the propagation direction of the LG beam \cite{19.1}.

\section{Cooperative Formation Probability Analysis}
To validate the feasibility of the proposed COW communication scheme, the cooperative formation probability $P_{{\textup{COW}}}$ is defined as the probability that BS can select at least two CUs as one CP. Furthermore, the cooperative formation probability $P_{{\textup{COW}}}$ is derived under the assumption that the users are randomly distributed in the coverage of BS. To this end, we first give the probability $P_\textup{s}(r_s)$ that two CUs locate at the same $r_{s}$ search ring. Then, based on the cosine theorem, we derive the probability ${P_{{\textup{d}}}}(r_s)$ that two CUs satisfy the communication distance requirement. Finally, the cooperative formation probability $P_{{\textup{COW}}}$ can be obtained by integrating the feasible search radius region.

The probability that one CU locates at the $r_{s}$ search ring can be derived as
\begin{align}\label{9abb}
{P_{{\textup{r}}}}(r_s)&{\rm{=}}\frac{{{S_{{\textup{r}}}}}}{{{S_{{\textup{BS}}}}}}{\rm{ = }} \frac{{{{\left( {r_{s} + \epsilon } \right)}^2} - {{\left( {r_{s} - \epsilon } \right)}^2}}}{{{R_{\textup {BS}}^2}}},
\end{align}
where ${S_{{\textup{r}}}}$ is the area of the $r_{s}$ search ring, ${S_{{\textup{BS}}}}$ represents the area of BS coverage, and ${R_{\textup {BS}}}$ is the BS coverage radius. Since two CUs should have the same distance to the BS, two CUs should locate at the same $r_{s}$ search ring. Furthermore, users are randomly distributed within the BS coverage, and the probability of two CUs located at the same $r_{s}$ search ring can be written as
\begin{equation}
P_\textup{s}(r_s)={P^2_{{\textup{r}}}(r_s)}.
\end{equation}

In the proposed COW communication scheme, device-to-device (D2D) communications are used to realize the COW communications. The distance between two CUs that locate at the same $r_{s}$ search ring should satisfy the communication distance condition $d_{{k_1},{k_2}} \in [ D_\textup{min}, D_\textup{max}]$. $D_\textup{max}$ is the maximum D2D communication distance that can guarantee the efficient information transmission.   Based on (\ref{11ee}) and (\ref{8abb}), the minimum communication distance $d_{{k_1},{k_2}}$ (i.e., $D_\textup{min}$) can be given as
\begin{align}\label{aaaaaa}
{D_{\min }}& =2R_\textup{fea}(z^c)\nonumber  \\
 & =2\sqrt {\frac{{\sqrt {{H^2} + {r_{s}^2} - {{\left( {\frac{{{D_{\min }}}}{2}} \right)}^2}}  \times \lambda  \times \left| l \right|}}{\pi }}.
\end{align}
Then, (\ref{aaaaaa}) can be transformed into the following expression
\begin{equation}
{D_\textup{min }}(r_s) =  \sqrt {\frac{{2\lambda \left| l \right|}}{\pi } \times\left( \sqrt {\frac{{{\lambda ^2}{l^2}}}{{{\pi ^2}}} + 4\left( {{H^2} + {r_{s}^2}} \right)}  - \frac{{{\lambda \left| l \right|}}}{{{\pi }}}\right)}.
\end{equation}
To obtain the probability that two CUs satisfy the communication distance condition $d_{{k_1},{k_2}} \in [ D_\textup{min}, D_\textup{max}]$, we calculate the minimum communication angle $\theta_\textup{min}(r_s)$ and maximum communication angle $\theta_\textup{max}(r_s)$ corresponding to the minimum communication distance $d_{{k_1},{k_2}}=D_\textup{min}$ and maximum communication distance $d_{{k_1},{k_2}}=D_\textup{max}$, respectively.

\begin{figure}[thp]
\setlength{\abovecaptionskip}{0.cm}
\setlength{\belowcaptionskip}{-0.cm}
\centering
\vspace{-0.35cm}
\setlength{\abovecaptionskip}{0.cm}
\setlength{\belowcaptionskip}{-0.cm}
\subfigure[]{\label{fig:bbb1}\includegraphics[height=1.7in,width=4.3cm]{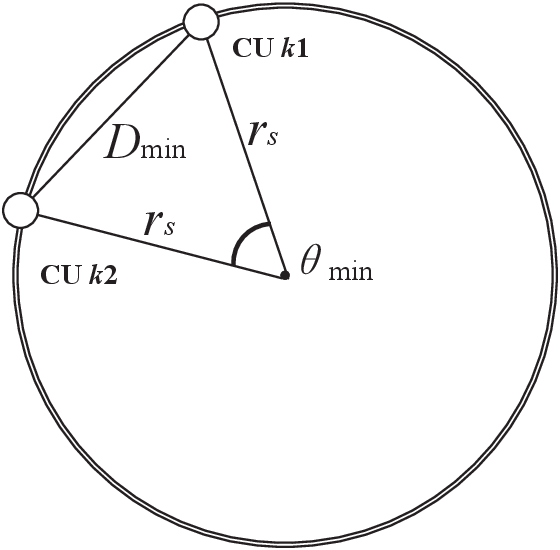}}
\subfigure[]{\label{fig:bbb2}\includegraphics[height=1.7in,width=4.3cm]{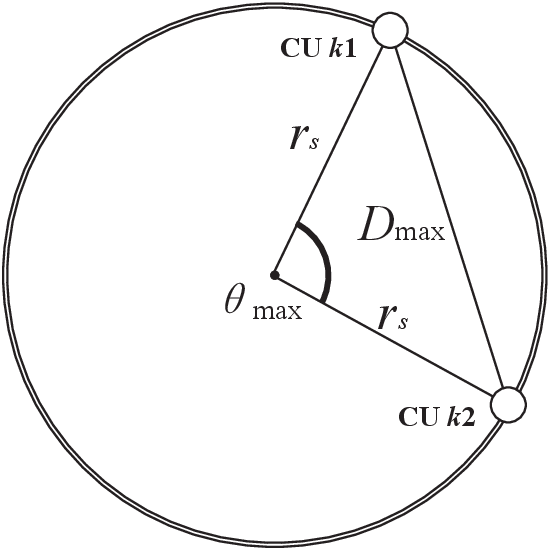}}
\caption{\!The minimum communication angle $\theta_\textup{min}(r_s)$ and maximum communication angle $\theta_\textup{max}(r_s)$}

\intextsep=1pt plus 3pt minus 1pt
\end{figure}
The minimum communication angle $\theta_\textup{min}(r_s)$ and maximum communication angle $\theta_\textup{max}(r_s)$ are shown in Figs. \ref{fig:bbb1} and \ref{fig:bbb2}, respectively. Based on the cosine theorem, the expressions of $\theta_\textup{min}(r_s)$ and $\theta_\textup{max}(r_s)$ can be respectively derived as
\begin{equation}
{\theta _\textup{min }}(r_s){\rm{ \triangleq }}{\arccos}\frac{{2{{{r_{s}}^2} - {D_\textup{min }^2}}}}{{2{{r_{s} }^2}}}
\end{equation}
and
\begin{equation}
{\theta _\textup{max }}(r_s){\rm{ \triangleq }}{\arccos}\frac{{2{{{r_{s}}^2} - {D_\textup{max }^2}}}}{{2{{r_{s} }^2}}}.
\end{equation}

We can observe that $\theta _\textup{max}(r_s)$ is equal to $\pi$ for $ r_s \in [{R_\textup{min }},\frac{{{D_\textup{max }}}}{2}]$,
where ${R_{{\textup{min}}}} $ can be expressed as follows:
\begin{equation}
{R_{{\textup{min}}}} = {R_{{\textup{fea}}}}(H) = \sqrt {\frac{{H \lambda  \left| l \right|}}{\pi }}.
\end{equation}
In general, the condition ${R_\textup{min }} \le \frac{{{D_\textup{max }}}}{2}$ holds due to the small value of wavelength $\lambda$. Therefore, the probability ${P_{{\textup{d}}}}(r_s)$ that two CUs satisfy the communication distance condition can be derived as
\begin{equation}
{P_{{\textup{d}}}}(r_s) = \frac{{\pi  - {\theta _\textup{min }}}}{\pi },
\end{equation}
which is also the proportion of the angle range that satisfies the communication distance condition.
Since $D_\textup{max }$ belongs to $[D_\textup{min}, {2R_\textup{BS}}]$, the maximum search radius is ${R_\textup{max }} = \min \left\{{    {R_{{\textup{v}}}},   {R_\textup{BS}}}  \right\}$
, where
\begin{equation}
{R_{{\textup{v}}}}{\rm{ = }}\sqrt {\left( {\frac{{{\pi ^2}D_\textup{max }^4}}{{16{\lambda ^2}{l ^2}}}} \right) + {{\left( {\frac{{{D_\textup{max }}}}{2}} \right)}^2} - {H^2}}.
\end{equation}
${R_{{\textup{v}}}}$ is obtained by solving the equation $D_\textup{min} = D_\textup{max}$.
For $\frac{{{D_\textup{max}}}}{2} < r_s \le {R_{{\textup{max}}}}$, the probability $P_{{\textup{d}}}(r_s)$ can be given as
\begin{equation}
{P_{{\textup{d}}}}(r_s) = \frac{{{\theta _\textup{max }}  - {\theta _\textup{min }}}}{\pi }.
\end{equation}
Based on the above analysis, the probability $P_{{\textup{d}}}(r_s)$ that two CUs satisfy the communication distance requirement can be summarized as follows:
\begin{equation}
{P_{{\textup{d}}}}(r_s) =
\left\{
        \begin{array}{cc}
        \frac{{\pi  - {\theta _\textup{min }}}}{\pi },              &{R_\textup{min }} \le r_s \le \frac{{{D_\textup{max }}}}{2}; \\
        \frac{{{\theta _\textup{max }} - {\theta _\textup{min }}}}{\pi }, &\frac{{{D_\textup{max }}}}{2} < r_s \le {R_{\textup{max}}};   \\
        0,                                                &{\rm{otherwise}}.
        \end{array}
\right.
\end{equation}

Then, we obtain the probability that two CUs locate at the same $r_{s}$ search ring and meet the communication distance condition, which can be expressed as
\begin{equation}
{P_{{\textup{c}}}}{\rm{ = }}\int_{0}^{{R_{{\textup{BS}} }}} {{P_{{\textup{s}}}(r_s)} \times {P_{{\textup{d}}}}(r_s)dr_s}.
\end{equation}
 $C_K^0$ represents the number of the combination that $0$ CU is chosen from $K$ users covered by the BS, which can be expressed as
\begin{equation}
C_K^0{\rm{ = }}\frac{{0!}}{{0!(K - 0)!}} = \frac{1}{{K!}}.
\end{equation}
Therefore, the cooperative formation probability is written as
\begin{equation}
{P_{{\textup{COW}}}}{\rm{ = 1}} - C_{{K}}^0 \times P_{{\textup{c}}}^0 \times {\left( {1 - {P_{{\textup{c}}}}} \right)^{{K}}},
\end{equation}
where $P_c^0$ denotes the zero power of $P_c$.

\section{Performance Analysis}
In the COW communications, there exists the BS, which can generate and transmit OAM signals. The two selected CUs, which are uniformly located at the $\pi$ angular arc, can sample OAM signals. For each CP, the phase-shift component is used to achieve 2 OAM mode multiplexing communications. Thus, the receive OAM signal can be given as
\begin{equation} \label{9}
{\textbf{y = }}{\textbf{G}{\textbf{Q}^{H}}\textbf{x} + \textbf{n}},
\end{equation}
where $\textbf{y} $ is the receive vector, $\textbf{G}$ represents the channel matrix, ${{\textbf{Q}}}^{\textbf{H}}$ denotes the multiplexing matrix, $\textbf{x}$ is the transmission information, and $\textbf{n}$ denotes the noise. ${{\textbf{Q}}^{\textbf{H}}}{\textbf{Q}}$ is the identity matrix. Based on (\ref{2}), the $(\ell,k)$-th element of the channel matrix $\textbf{G}$ can be expressed as
\begin{equation}
g(\ell,k)=U_\ell(r_k,z^{\rm{c}})=\frac {2}{\pi w^{2}(z^{\rm{c}})|\ell |!} \left [{\frac {\sqrt {2}r_k}{w(z^{\rm{c}})}}\right]^{2|\ell |} \!\!\!\!\!\!\exp \left [{\frac {-2r_k^{2}}{w^{2}(z^{\rm{c}})}}\right].
\end{equation}
For the COW communications, the channel matrix $\textbf{G}$ can be further written as
\begin{equation}\label{11}
\textbf{G} = {\textbf{Q}^{H}}\Lambda{\textbf{Q}},
\end{equation}
where $\Lambda  = \rm{diag}\{ {\beta _1},{\beta _2}\} $ is a diagonal matrix with the eigenvalues of $\textbf{G}$ on its diagonal. The $\ell$-th diagonal element $\beta_\ell$ is the transmission gain corresponding to the $\ell$-th OAM channel. Note that we only consider the OAM mode $\ell={1,2}$, which can obtain the highest spectrum efficiency. For the $\ell$-mode OAM transmission, the demultiplexing OAM signal can be written as
\begin{equation}
y_{\rm{de}}(\ell ) = {q(\ell )}({\textbf{G}{q^{H}(\ell )}x + \textbf{n}}) = {g_{{\rm{de}}}}(\ell )x + \tilde n,
\end{equation}
where $q(\ell ) \in {\textbf{Q}}$, ${g_{{\rm{de}}}}(\ell ) = q(\ell )\textbf{G}{q^{H}(\ell )}$ is the $\ell$-mode effective OAM channel and $\tilde n = q(\ell ) n $. For the COW communications, the spectrum efficiency of $\ell$-mode OAM communication can be given as follows:
\begin{equation}
C(\ell)= \textup{log}_2\left (1+\frac{P_\ell{\left| {{g_{{\rm{de}}}}(\ell )} \right|^2}}{\sigma^2}\right ),
\end{equation}
where $P_\ell$ denotes the power allocated to the $\ell$-th OAM channel based on the water-filling method and $\sigma^2$ is the noise power.

For the proposed COW communication scheme, selected CUs are required to locate at the feasible CU region, which is the limited angular range on the same circumference. To expand the feasible CU region,  OAM beam steering can be used to compensate the phase distortion based on oblique angles $\phi$ and $\psi$. The oblique angle $\phi$ is the included angle between the vertical line of the connection between the two CUs and the connection between the UCA center of BS and the midpoint of the two CUs. The oblique angle $\psi$ is the included angle between the vertical line of the UCA and the connection between the UCA center of BS and the midpoint of the two CUs.

Based on \cite{15a}, the transmit beam steering vector and receive beam steering vector are written as $\textbf{b}  =  [1, {e^{-j{{V_2}}}},..., {e^{-j{{V_M}}}}]$ and $\textbf{a}  =  [1, {e^{-j{{T_2}}}},..., {e^{-j{{T_{K_{\rm{c}}}}}}}]$, respectively. Furthermore,
\begin{equation}
V_m = \frac{{2\pi R}}{\lambda }\cos (\frac{{2\pi(m-1) }}{M} - \frac{\pi }{2})\sin \psi
\end{equation}
\begin{equation}
T_{k} = \frac{{2\pi {r_{{\rm{max}}}}({z^{\rm{c}}})}}{\lambda }\cos (\frac{{2\pi(k-1) }}{K_{\rm{c}}} + \frac{\pi }{2})\sin \phi ,
\end{equation}
where $m$ denotes the $m$-th transmit antenna element, $k$ is the $k$-th CU, and $K_{\rm{c}}$ represents the number of selected CUs.
By utilizing the beam steering, the $\ell$-mode effective OAM channel can be rewritten as follows:
\begin{equation}
{g'_{{\rm{de}}}}(\ell ) = (q(\ell ) \odot {\textbf{a}}){\textbf{G}}(\textbf{b}^{T} \odot {q^H}(\ell )).
\end{equation}
For the COW communications with beam steering, the spectrum efficiency of $\ell$-mode OAM communication can be expressed as
\begin{equation}
C'(\ell)=\textup{log}_2\left (1+\frac{P_\ell{\left| {{g'_{{\rm{de}}}}(\ell )} \right|^2}}{\sigma^2}\right ).
\end{equation}

Based on the above analysis, the expansion of the feasible CU region can be achieved through the compensation of the phase distortion based on oblique
angles by OAM beam steering, which can relax the limited angular range on the same circumference for selected CUs.
This enlarges the $\epsilon$ in (\ref{9abb}), which can be expressed as
\begin{equation}\label{9abbcc}
{\epsilon \rm{ = }}\frac{{{D_{\max }}\sin (\phi+\psi) }}{2}.
\end{equation}
As $\epsilon$ becomes larger, the cooperative formation probability $P_{{\textup{COW}}}$ will become larger, which makes our proposed COW communication scheme more feasible.

\section{Simulation Results}

In this section, simulation results are provided to evaluate the performance of our proposed COW communication scheme. The users are randomly distributed in the BS coverage. We set the operating frequency, BS transmit power, and noise power as $1~\textup{GHz}$, $P=30~\textup{dBm}$, and $\sigma^2=-90~\textup{dBm}$, respectively. We compare the proposed COW communication scheme with the traditional OAM wireless communications with the fixed UCA scheme.

\begin{figure}[htb]
\setlength{\abovecaptionskip}{0.cm}
\setlength{\belowcaptionskip}{-0.cm}
\centering
\vspace{-0.35cm}
\setlength{\abovecaptionskip}{0.cm}
\setlength{\belowcaptionskip}{-0.cm}
\includegraphics[height=2.8in,width=8.5cm]{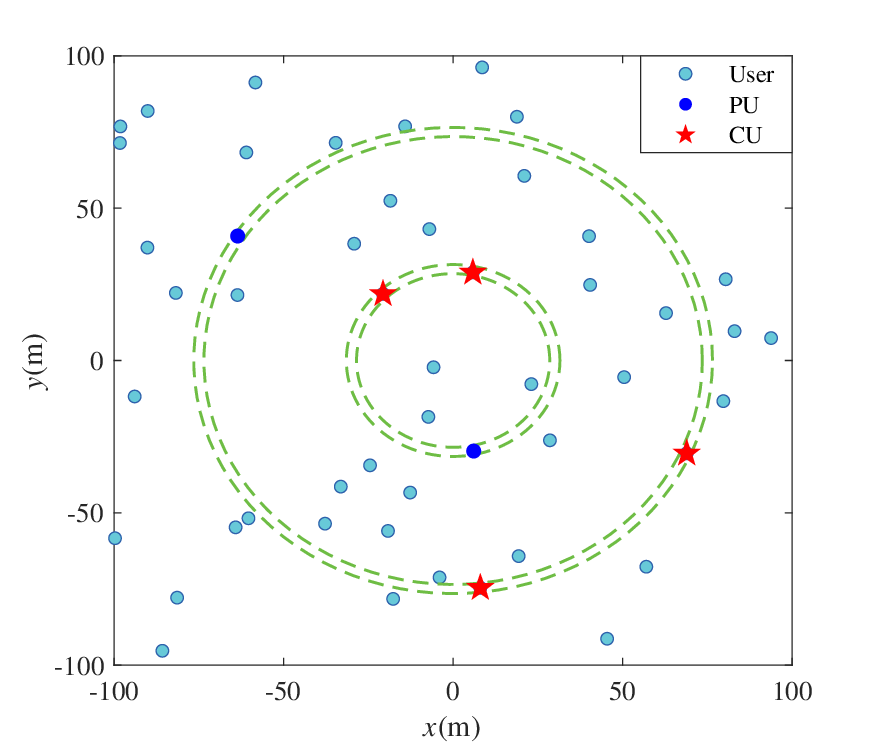}
\caption{\!The illustration of the selected CUs}
\label{fig:a1}
\intextsep=1pt plus 3pt minus 1pt
\end{figure}

Figure \ref{fig:a1} is given to demonstrate the selected CUs by using our proposed COW communication scheme. 50 users marked by ``$\circ$" are uniformly distributed in $200~\textup{m}\times200~\textup{m}$ area, and the green dashed rings are the determined $r_{s}$ search rings. 6 users in the green dashed rings are the PUs, and the ``\ding{73}" are the selected CUs, which can achieve the COW communications.

\begin{figure}[htb]
\setlength{\abovecaptionskip}{0.cm}
\setlength{\belowcaptionskip}{-0.cm}
\centering
\vspace{-0.25cm}
\setlength{\abovecaptionskip}{0.cm}
\setlength{\belowcaptionskip}{-0.cm}
\includegraphics[height=2.8in,width=8.7cm]{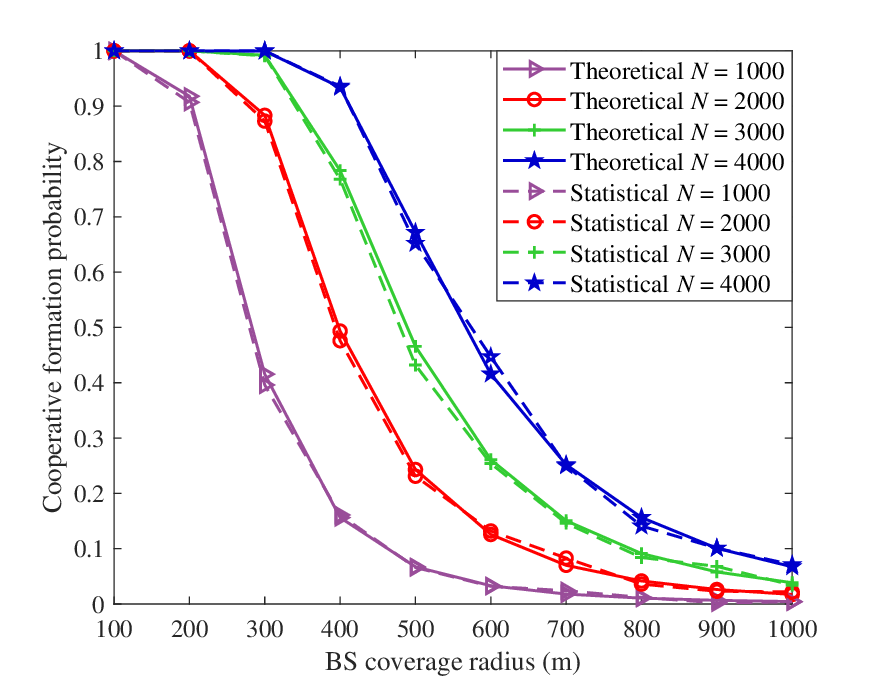}
\caption{\!The cooperative formation probability versus BS coverage radius}
\label{fig:a2}
\intextsep=1pt plus 3pt minus 1pt
\end{figure}
For the proposed COW communication scheme, Fig. \ref{fig:a2} plots the cooperative formation probability versus BS coverage radius with different user numbers 1000, 2000, 3000, and 4000. It can be observed that the cooperative formation probability is a decreasing function of the BS coverage radius. The proposed COW communication scheme with user number 4000 obtains the largest cooperative formation probability. This is due to the fact that the increase of user density (i.e., user number in the BS coverage) can decrease the distance between users, which increases the probability to select at least two CUs as one CP. Furthermore, we can see that the theoretical cooperative formation probability is consistent with the statistical cooperative formation probability, which validates the correctness of our derived closed-form theoretical cooperative formation probability.

\begin{figure}[htb]
\setlength{\abovecaptionskip}{0.cm}
\setlength{\belowcaptionskip}{-0.cm}
\centering
\vspace{-0.25cm}
\setlength{\abovecaptionskip}{0.cm}
\setlength{\belowcaptionskip}{-0.cm}
\includegraphics[height=2.8in,width=8.7cm]{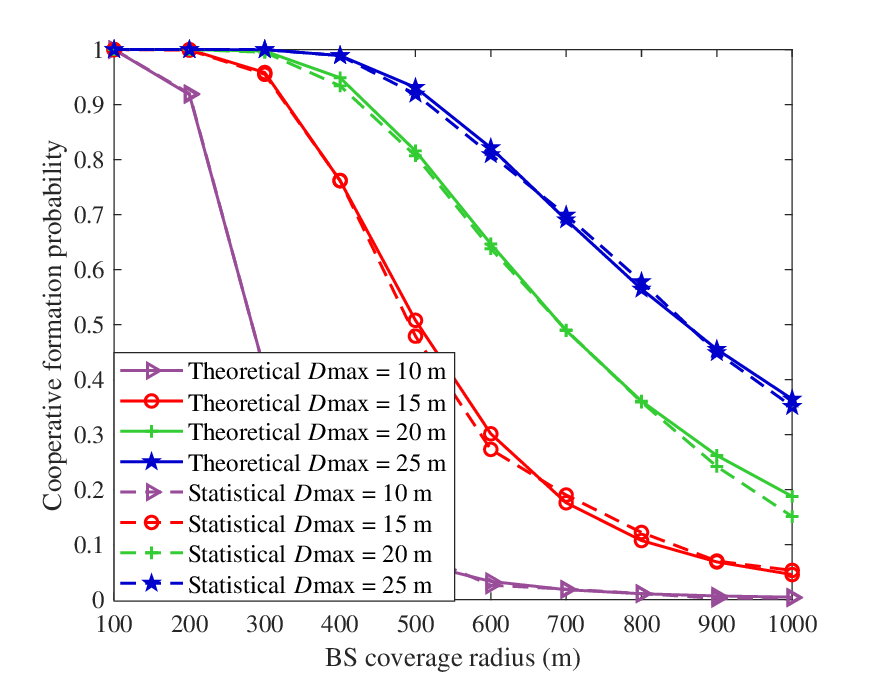}
\caption{\!The cooperative formation probabilities with different maximum transmission distances between two CUs}
\label{fig:b}
\intextsep=1pt plus 3pt minus 1pt
\end{figure}
 Under various BS coverage radii, the cooperative formation probabilities with different maximum transmission distances 10 m, 15 m, 20 m, and 25 m are illustrated in Fig. \ref{fig:b} for the validation of the proposed COW communication scheme. For all the maximum transmission distances between two CUs, the cooperative formation probabilities monotonically decrease as the BS coverage radius increases. In addition, we can see that the larger the maximum transmission distance between two CUs is, the larger the cooperative formation probability is. This is because the increase of the maximum transmission distance between two CUs can increase the number of PUs that can satisfy the communication distance condition of the proposed COW communication scheme. In this paper, we set the maximum transmission distance between two CUs as 10 m, which can guarantee the LoS path between two CUs and achieve efficient cooperative communications for two CUs.

\begin{figure}[htb]
\setlength{\abovecaptionskip}{0.cm}
\setlength{\belowcaptionskip}{-0.cm}
\centering
\vspace{-0.25cm}
\setlength{\abovecaptionskip}{0.cm}
\setlength{\belowcaptionskip}{-0.cm}
\includegraphics[height=2.8in,width=8.7cm]{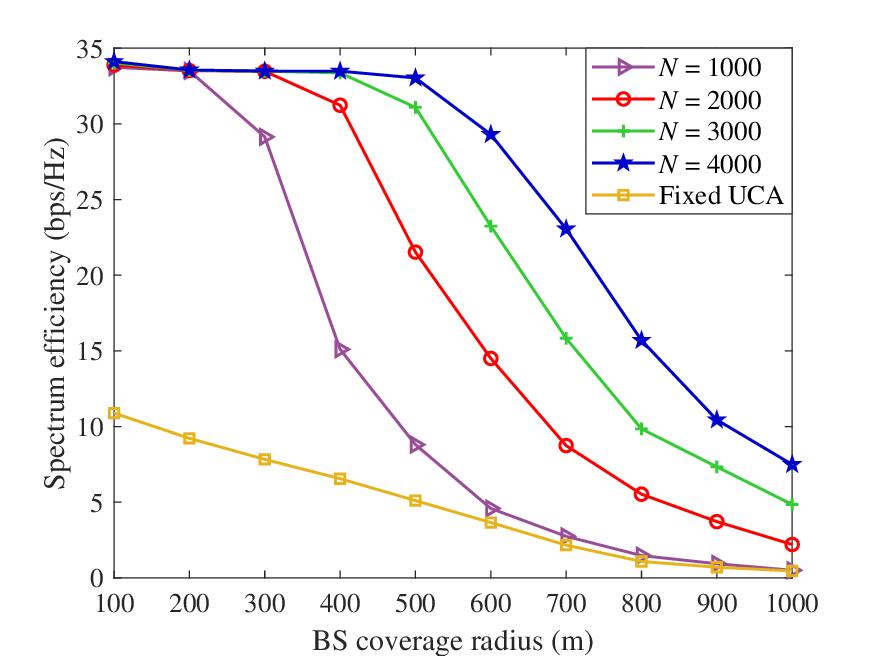}
\caption{\!The spectrum efficiency of the proposed COW communication scheme}
\label{fig:c1e}
\intextsep=1pt plus 3pt minus 1pt
\end{figure}

Figure \ref{fig:c1e} shows the spectrum efficiency of the proposed COW communication scheme with different user numbers.
For the fixed UCA scheme, there are two antennas to receive the OAM signal.
It can be seen that the proposed COW communication scheme obtains higher spectrum efficiency than the fixed UCA scheme.
We can observe that the spectrum efficiency is a monotonic decreasing function of the BS coverage radius.
Furthermore, the larger the user density is, the larger the spectrum efficiency of the proposed COW communication scheme is. This is due to the fact that with the increase of the user density for the fixed BS coverage radius, the proposed COW communication scheme has the larger probability to select the CUs that can efficiently receive the OAM signal. Therefore, the proposed COW communication scheme can support the high spectrum efficiency over the long transmission distance. The more CUs are, the higher the spectrum efficiency of the proposed COW communication scheme is.

\begin{figure}[tb]
\setlength{\abovecaptionskip}{0.cm}
\setlength{\belowcaptionskip}{-0.cm}
\centering
\vspace{-0.25cm}
\setlength{\abovecaptionskip}{0.cm}
\setlength{\belowcaptionskip}{-0.cm}
\includegraphics[height=2.8in,width=8.7cm]{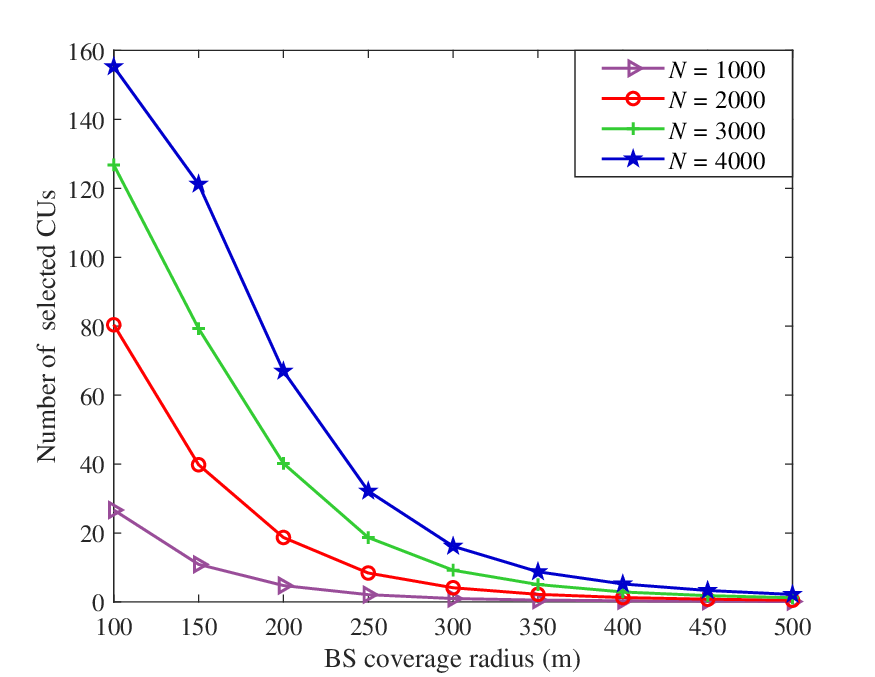}
\caption{\!The number of selected CUs}
\label{fig:d}
\intextsep=1pt plus 3pt minus 1pt
\end{figure}

 In Fig. \ref{fig:d}, the number of CUs selected by the proposed COW communication scheme is depicted under various BS coverage radii. It can be observed that the number of selected CUs monotonically decreases as the BS coverage radius increases. Moreover, as the user density increases, the proposed COW communication scheme can select more CUs to achieve COW communications for spectrum efficiency improvement. This is because the proposed COW communication scheme has the large probability to select the CUs for the large user density.

\begin{figure}[tb]
\setlength{\abovecaptionskip}{0.cm}
\setlength{\belowcaptionskip}{-0.cm}
\centering
\vspace{-0.25cm}
\setlength{\abovecaptionskip}{0.cm}
\setlength{\belowcaptionskip}{-0.cm}
\includegraphics[height=2.8in,width=8.7cm]{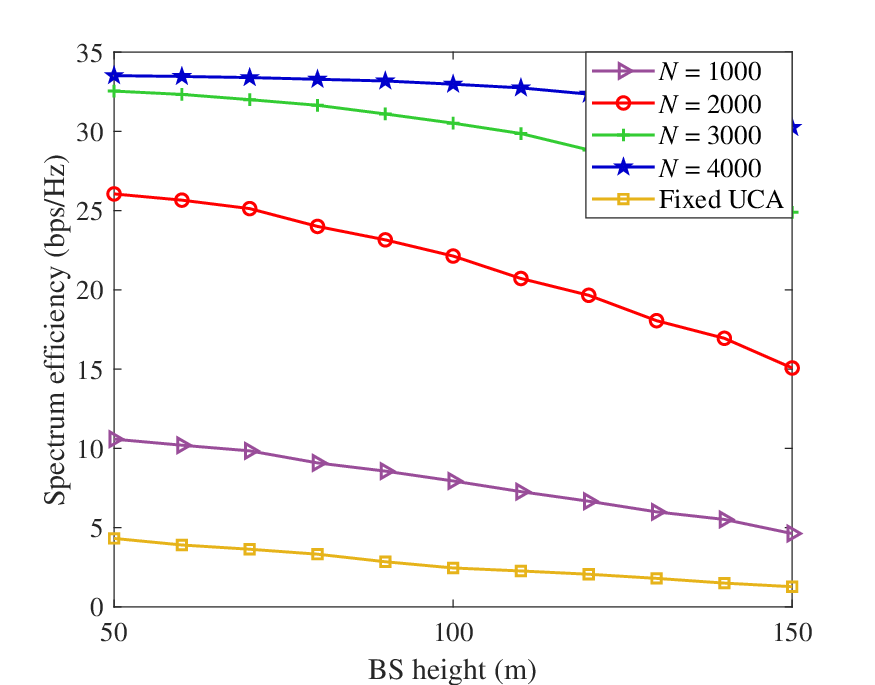}
\caption{\!The spectrum efficiency versus BS height}
\label{fig:c2e}
\intextsep=1pt plus 3pt minus 1pt
\end{figure}

 Figure \ref{fig:c2e} presents the spectrum efficiency performance under various BS heights.
There are two antennas to receive the OAM signal for the fixed UCA scheme.
The proposed COW communication scheme achieves better spectrum efficiency performance than the fixed UCA scheme.
From Fig. \ref{fig:c2e}, we can see that the proposed COW communication scheme with user number 4000 can obtain higher spectrum efficiency than those with user numbers 1000, 2000, and 3000. Furthermore, the spectrum efficiency of the proposed COW communication scheme is a monotonic decreasing function of BS height. This is due to the fact that as the BS height increases, the transmission distance of the OAM signal will increase, thus making the minimum communication distance between two CUs larger. Based on the proposed COW communication scheme, it can be observed that as the minimum communication distance between the two CUs increases, the proposed COW communication scheme has the small probability to select the CUs.

\begin{figure}[tb]
\setlength{\abovecaptionskip}{0.cm}
\setlength{\belowcaptionskip}{-0.cm}
\centering
\vspace{-0.25cm}
\setlength{\abovecaptionskip}{0.cm}
\setlength{\belowcaptionskip}{-0.cm}
\includegraphics[height=2.8in,width=8.7cm]{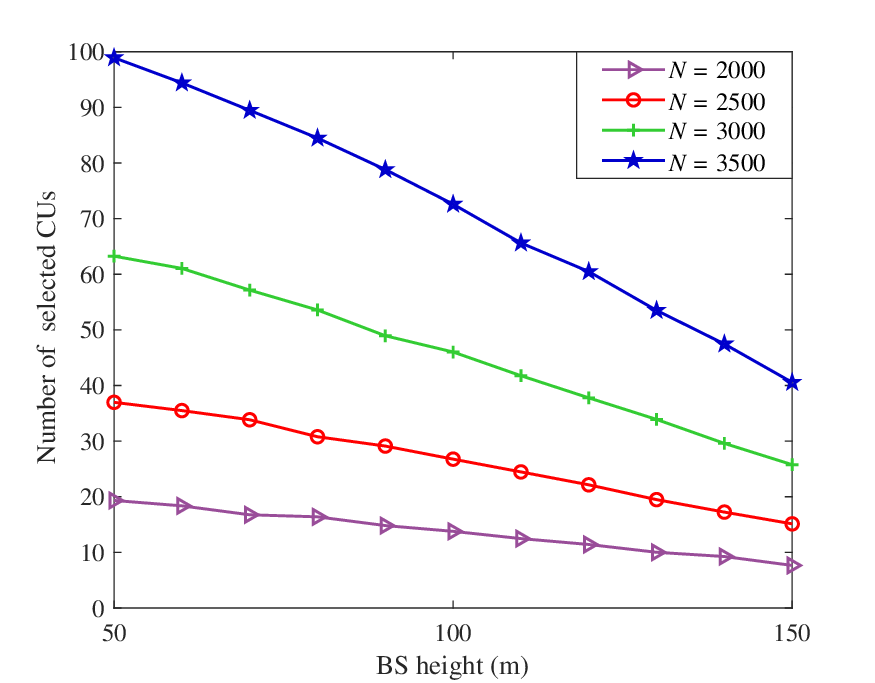}
\caption{\!The number of selected CUs versus BS height}
\label{fig:f}
\intextsep=1pt plus 3pt minus 1pt
\end{figure}

Figure \ref{fig:f} demonstrates the number of selected CUs versus BS height. We can observe that the number of CUs selected by the proposed COW communication scheme monotonically decreases as the BS height increases. Moreover, with the increase of the user density, the proposed COW communication scheme can select more CUs. This is because as the user density increases, the proposed COW communication scheme obtains the large probability to realize long distance information transmission. The larger the user density is, the more the number of selected CUs is.

\begin{figure}
\setlength{\abovecaptionskip}{0.cm}
\setlength{\belowcaptionskip}{-0.cm}
\centering
\vspace{-0.25cm}
\setlength{\abovecaptionskip}{0.cm}
\setlength{\belowcaptionskip}{-0.cm}
\includegraphics[height=2.8in,width=8.7cm]{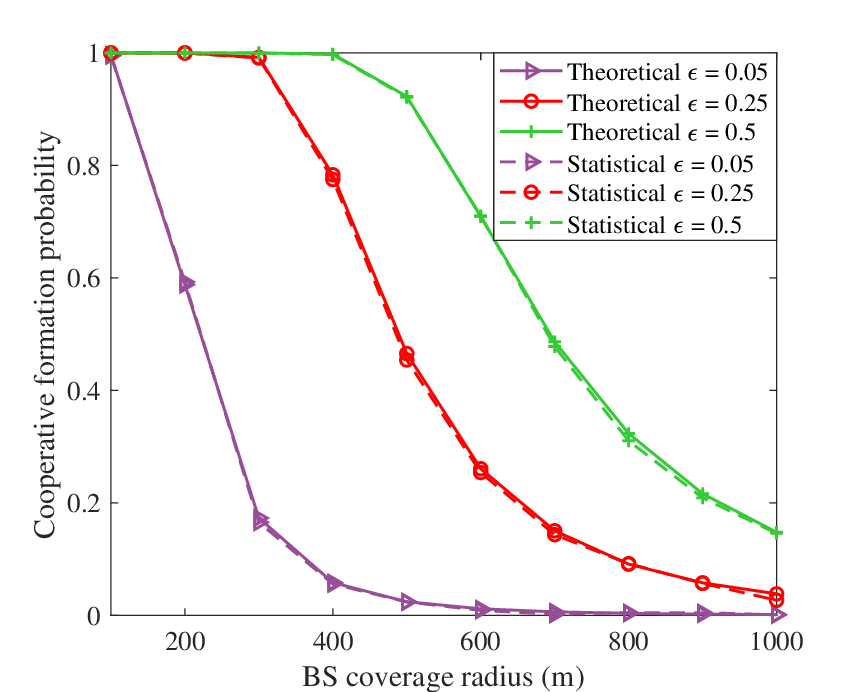}
\caption{\!The cooperative formation probability versus BS coverage radius}
\label{fig:a1bb}
\intextsep=1pt plus 3pt minus 1pt
\end{figure}

For the proposed COW communication scheme, Fig. \ref{fig:a1bb} plots the cooperative formation probability with different $\epsilon$. The maximum transmission distance is set as $D_{\max}=10$m, and the oblique angle is $6^{\circ}$ based on \cite{15a}. According to (\ref{9abbcc}), we can obtain $\epsilon\approx 0.52$. Therefore, the cooperative formation probability is plotted with $\epsilon=0.05,~0.25,~\rm{and}~0.5$. It can be observed that the larger the $\epsilon$ is, the larger the cooperative formation probability is. This is due to the fact that the increase of $\epsilon$ can expand the feasible CU region, which increases the number of PUs and makes our proposed COW communication scheme more feasible. The cooperative formation probability monotonically decreases as the BS coverage radius increases.

\section{Conclusions}

In this paper, a COW communication scheme, which realizes the effective reception of the OAM signal, is proposed to select the CUs with size-limited equipment for COW communications at the long distance. To this end, the feasible radial radius and selective waist radius were obtained by deriving the intensity distribution with respect to the radial radius. Then, the two CUs were chosen by using the same circle with the origin at the BS. Finally, to guarantee the reception of the OAM signal, we adjusted the waist radius to make the two CUs at the maximum intensity. Simulation results showed that compared with the traditional OAM wireless communications with the fixed UCA scheme, the proposed COW communication scheme can achieve higher spectrum efficiency over the long transmission distance. Furthermore, the theoretical cooperative formation probability was very close to the statistical cooperative formation probability, which validated the feasibility of the proposed COW communication scheme.

\begin{IEEEbiography}[{\includegraphics[width=0.9in,height=1.25in]{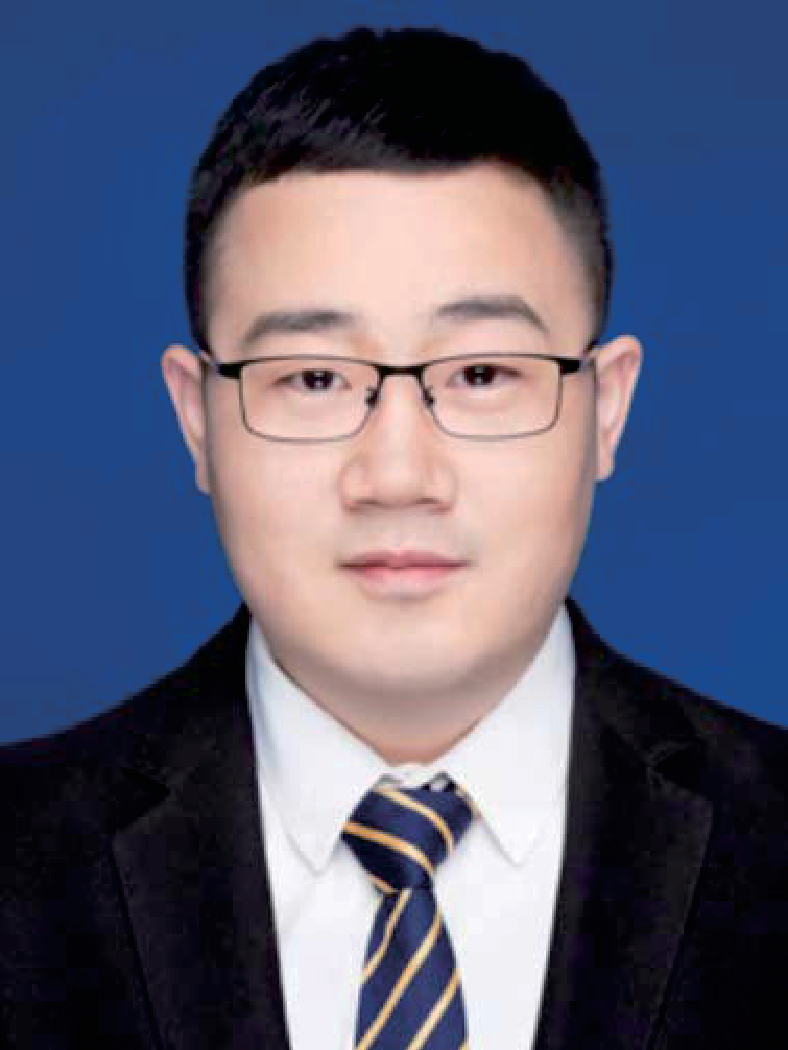}}]{Ruirui Chen}
received the Ph.D. degree in military communication from Xidian University, Xi'an, China, in 2018.

He is currently a lecturer with the School of Information and Control Engineering, China University of Mining and Technology, Xuzhou, China. He is also a master supervisor of electronic information in China University of Mining and Technology. His research interests focus on OAM wireless communications, UAV communications and B5G/6G wireless communications.
\end{IEEEbiography}

\begin{IEEEbiography}[{\includegraphics[width=0.9in,height=1.25in]{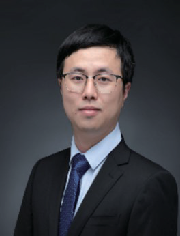}}]{Wenchi Cheng}
(Senior Member, IEEE) received the B.S. and Ph.D. degrees in telecommunication engineering from Xidian University, Xi'an, China, in 2008 and 2013, respectively.
He was a Visiting Scholar with the Department of Electrical and Computer Engineering, Texas A\&M University, College Station, TX, USA, from 2010 to 2011. He is currently a Full Professor with Xidian University. He has published more than 100 international journal and conference papers in IEEE JOURNAL ON SELECTED AREAS IN COMMUNICATIONS, IEEE Magazines, IEEE TRANSACTIONS, IEEE INFOCOM, IEEE GLOBECOM, and IEEE ICC. His current research interests include B5G/6G wireless networks, emergency wireless communications, and OAM wireless communications.

Prof. Cheng received the IEEE ComSoc Asia-Pacific Outstanding Young Researcher Award in 2021, the URSI Young Scientist Award in 2019, the Young Elite Scientist Award of CAST, and four IEEE journal/conference best papers award. He has served or is serving as an Associate Editor for IEEE SYSTEMS JOURNAL, IEEE COMMUNICATIONS LETTERS, and IEEE WIRELESS COMMUNICATIONS LETTERS; the Wireless Communications Symposium Co-Chair for IEEE ICC 2022 and IEEE GLOBECOM 2020; the Publicity Chair for IEEE ICC 2019; the Next Generation Networks Symposium Chair for IEEE ICCC 2019; and the Workshop Chair for IEEE ICC 2019/IEEE GLOBECOM 2019/INFOCOM 2020 Workshop on Intelligent Wireless Emergency Communications Networks.
\end{IEEEbiography}
\vspace{-100 mm}
\begin{IEEEbiography}[{\includegraphics[width=0.9in,height=1.25in]{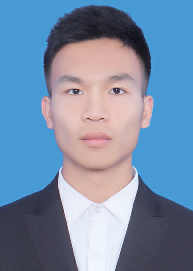}}]{Jinyang Lin}
received the B.S. degree in information and communication engineering from China University of Mining and Technology, Xuzhou, China, in 2021. He is currently working towards the M.S. degree with the School of Information and Control Engineering, China University of Mining and Technology, Xuzhou, China. His current research interests focus on in OAM wireless communications. Email: linjinyang@cumt.edu.cn.
\end{IEEEbiography}
\vspace{-100 mm}
\begin{IEEEbiography}[{\includegraphics[width=0.9in,height=1.25in]{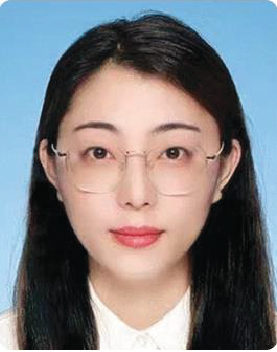}}]{Liping Liang}
received the Ph.D. degree in communication and information system with Xidian University, Xi'an, China. She is a prospective associate professor with Xidian University. Her research interests focus on 5G wireless communications with emphasis on radio OAM wireless communications and anti-jamming communications. Email: liangliping@xidian.edu.cn.
\end{IEEEbiography}


\begin{thebibliography}{99}
\bibitem{1}
R. Lyu, W. Cheng, B. Shen, et al., ``OAM-SWIPT for IoE-Driven 6G," \emph{IEEE Communications Magazine}, vol. 60, no. 3, pp. 19-25, Mar. 2022.

\bibitem{2}
R. Chen, H. Zhou, M. Moretti, et al., ``Orbital angular momentum waves: generation, detection, and emerging applications," \emph{IEEE Communications Surveys and Tutorials}, vol. 22, no. 2, pp. 840-868, 2020.

\bibitem{3}
S. M. Mohammadi, L. K. S. Daldorff, J. E. S. Bergman, et al., ``Orbital angular momentum in radio--a system study," \emph{IEEE Transactions on Antennas and Propagation}, vol. 58, no. 2, pp. 565-572, Feb. 2010.

\bibitem{4}
R. Chen, F. Cheng, J. Lin, et al., ``Performance analysis of rate splitting multiple access based vortex wave communications," \emph{IEEE Wireless Communications Letters}, vol. 11, no. 8, pp. 1570-1574, Aug. 2022.

\bibitem{5}
R. Lyu, W. Cheng and W. Zhang, ``Modeling and performance analysis of OAM-NFC systems," \emph{IEEE Transactions on Communications}, vol. 69, no. 12, pp. 7986-8001, Dec. 2021.

\bibitem{7}
W. Cheng, W. Zhang, H. Jing, et al., ``Orbital angular momentum for wireless communications," \emph{IEEE Wireless Communications}, vol. 26, no. 1, pp. 100-107, Feb. 2019.

\bibitem{7a}
L. Liang, W. Cheng, W. Zhang, et al., ``Mode hopping for anti-jamming in radio vortex wireless communications," \emph{IEEE Transactions on Vehicular Technology}, vol. 67, no. 8, pp. 7018-7032, Aug. 2018.

\bibitem{6}
T. Hu, Y. Wang, X. Liao, et al., ``OAM-based beam selection for indoor millimeter wave MU-MIMO systems," \emph{IEEE Communications Letters}, vol. 25, no. 5, pp. 1702-1706, May 2021.

\bibitem{8}
A. A. Amin and S. Y. Shin, ``Capacity analysis of cooperative NOMA-OAM-MIMO based full-duplex relaying for 6G," \emph{IEEE Wireless Communications Letters}, vol. 10, no. 7, pp. 1395-1399, Jul. 2021.


\bibitem{10}
L. Allen, M. W. Beijersbergen, R. J. C. Spreeuw, et al., ``Orbital angular momentum of light and the transformation of LaguerreGaussian laser modes," \emph{Physical Review A}, vol. 45, no. 11, pp. 8185-8189, Jun. 1992.

\bibitem{10A}
L. Zhu, H. Yao, H. Chang, et al.,``Adaptive optics for orbital angular momentum-based internet of underwater things applications," \emph{IEEE Internet of Things Journal}, vol. 9, no. 23, pp. 24281-24299, Dec. 2022.

\bibitem{10A1}
J. Li, Q. Yang, X. Dai, et al., ``Joint beam-and-probabilistic shaping scheme based on orbital angular momentum mode for indoor optical wireless communications," \emph{Journal of Lightwave Technology}, 2023. Available: https://ieeexplore.ieee.org/stamp/stamp.jsp?t p=\&arnumber=10158487.

\bibitem{10A2}
Y. Ren, L. Li, G. Xie, et al., ``Line-of-sight millimeter-wave communications using orbital angular momentum multiplexing combined with conventional spatial multiplexing," \emph{IEEE Transactions on Wireless Communications}, vol. 16, no. 5, pp. 3151-3161, May 2017.

 \bibitem{10B}
H. Jing, W. Cheng, W. Zhang, et al., ``Optimal UCA design for OAM based wireless backhaul transmission," \emph{IEEE International Conference on Communications}, pp. 1-6, 2020.

\bibitem{10C}
Z. Zhu, C. Zhang, L. Wang, et al., ``Research on orbital angular momentum different modes networking method in wireless communication," \emph{IEEE Wireless Communications Letters}, vol. 11, no. 5, pp. 1007-1011, May 2022.

\bibitem{10a}
K. A. Opare, Y. Kuang, and J. J. Kponyo, ``Mode combination in an ideal wireless OAM-MIMO multiplexing system," \emph{IEEE Wireless Communications Letters}, vol. 4, no. 4, pp. 449-452, Aug. 2015.

\bibitem{10b}
W. Cheng, H. Zhang, L. Liang, et al., ``Orbital angular-momentum embedded massive MIMO: achieving multiplicative spectrum-efficiency for mmWave communications," \emph{IEEE Access}, vol. 6, pp. 2732-2745, 2018.


\bibitem{11a}
F. E. Mahmouli and S. D. Walker, ``4-Gbps uncompressed video transmission over a 60-GHz orbital angular momentum wireless channel," \emph{IEEE Wireless Communications Letters}, vol. 2, no. 2, pp. 223-226, 2013.

\bibitem{11}
F. Tamburini, E. Mari, A. Sponselli, et al., ``Encoding many channels in the same frequency through radio vorticity: first experimental test," \emph{New Journal of Physics}, vol. 14, no. 3, Art. 2012.

\bibitem{12}
W. Zhang, S. Zheng, X. Hui, et al., ``Mode division multiplexing communication using microwave orbital angular momentum: an experimental study," \emph{IEEE Transactions on Wireless Communications}, vol. 16, no. 2, pp. 1308-1318, Feb. 2017.

\bibitem{12a}
Y. Zhao and C. Zhang, ``Compound angular lens for radio orbital angular momentum coaxial separation and convergence," \emph{IEEE Antennas and Wireless Propagation Letters}, vol. 18, no. 10, pp. 2160-2164, 2019.

\bibitem{13}
L. Liang, W. Cheng, W. Zhang, et al., ``Joint OAM multiplexing and OFDM in sparse multipath environments," \emph{IEEE Transactions on Vehicular Technology}, vol. 69, no. 4, pp. 3864-3878, Apr. 2020.

\bibitem{14}
D. Lee, H. Sasaki, H. Fukumoto, et al., ``An experimental demonstration of 28 GHz band wireless OAM-MIMO (orbital angular momentum multi-input and multi-output) multiplexing," \emph{2018 IEEE 87th Vehicular Technology Conference (VTC Spring)}, pp. 1-5, 2018.
\bibitem{14a}
X. Xiong, S. Zheng, Y. Chen, et al., ``Plane spiral OAM mode-group orthogonal multiplexing communication using partial arc sampling receiving scheme," \emph{IEEE Transactions on Antennas and Propagation}, vol. 70, no. 11, pp. 10998-11008, Nov. 2022.
\bibitem{14b}
X. He, L. Deng, B. Feng, et al., ``Angular momentum multiplexing via a shared-aperture patch antenna," \emph{IEEE Transactions on Antennas and Propagation}, 2023. Available: https://ieeexpl-ore.ieee.org/stamp/stamp.jsp?tp=\&arnumber=10024147.

\bibitem{15}
Z. Zhang, Y. Xiao, Z. Ma, et al., ``6G wireless networks: vision, requirements, architecture, and key technologies," \emph{IEEE Vehicular Technology Magazine}, vol. 14, no. 3, pp. 28-41, Sept. 2019.

\bibitem{15a}
R. Chen, H. Xu, M. Moretti, et al., ``Beam steering for the misalignment in UCA-based OAM communication systems," \emph{IEEE Wireless Communications Letters}, vol. 7, no. 4, pp. 582-585, 2018.

\bibitem{15b}
H. Jing, W. Cheng and X. Xia, ``A simple channel independent beamforming scheme with parallel uniform circular array," \emph{IEEE Communications Letters}, vol. 23, no. 3, pp. 414-417, Mar. 2019.

\bibitem{15c}
K. Liu, Y. Cheng, Z. Yang, et al., ``Orbital angular-momentum-based electromagnetic vortex imaging," \emph{IEEE Antennas and Wireless Propagation Letters}, vol. 14, pp. 711-714, 2015.


\bibitem{16}
P. Yang, Y. Xiao, M. Xiao, et al., ``6G wireless communications: vision and potential techniques," \emph{IEEE Network}, vol. 33, no. 4, pp. 70-75, Aug. 2019.

\bibitem{16a}
Y. Zhang, W. Feng, and N. Ge, ``On the restriction of utilizing orbital angular momentum in radio communications," \emph{2013 8th International Conference on Communications and Networking in China (CHINACOM)}, pp. 271-275, Aug. 2013.

\bibitem{17}
G. Xie, L. Li, Y. Ren, et al., ``Performance metrics and design considerations for a free-space optical orbital-angular-momentum-multiplexed communication link," \emph{Optica}, vol. 2, no. 4, pp. 357-365, 2015.

\bibitem{17a}
Q. Song, Y. Wang, K. Liu, et al., ``Beam steering for OAM beams using time modulated circular arrays," \emph{Electronics Letters}, vol. 54, no. 17, pp. 1017-1018, 2018.

\bibitem{18}
Z. Tian, R. Chen, W. Long, et al., ``Broadband beam steering for misaligned multi-mode OAM communication systems," \emph{Journal of Systems Engineering and Electronics}, vol. 32, no. 4, pp. 779-788, Aug. 2021.

\bibitem{19}
R. Chen, W. Long, X. Wang, et al., ``Multi-mode OAM radio waves: generation, angle of arrival estimation and reception with UCAs," \emph{IEEE Transactions on Wireless Communications}, vol. 19, no. 10, pp. 6932-6947, Oct. 2020.

\bibitem{19.1}
R. Chen, Z. Tian, W. Long, et al., ``Hybrid mechanical and electronic beam steering for maximizing OAM channel capacity," \emph{ IEEE Transactions on Wireless Communications}, vol. 22, no. 1, pp. 534-549, Jan. 2023.

 \bibitem{19A1}
M. J Padgett, F. M Miatto, M. P J Lavery, et al.,``Divergence of an orbital-angular-momentum-carrying beam upon propagation," \emph{ New Journal of Physics}, vol.17, no.2, 2015.

\bibitem{19A2}
M. Oldoni, F. Spinello, E. Mari, et al.,``Space-division demultiplexing in orbital-angular-momentum-based MIMO radio systems," \emph{ IEEE Transactions on Antennas and Propagation}, vol. 63, no. 10, pp. 4582-4587, Oct. 2015.

\bibitem{19B}
G. Xie, Y. Ren, H. Huang, et al., ``Analysis of aperture size for partially receiving and de-multiplexing 100-gbit/s optical orbital angular momentum channels over free-space link," \emph{2013 IEEE Globecom Workshops (GC Wkshps)}, pp. 1116-1120, Dec. 2013.
\bibitem{19.2}
R. Chen, J. Zhou, W. Long, et al., ``Hybrid circular array and luneberg lens for long-distance OAM wireless communications," \emph{IEEE Transactions on Communications}, vol. 71, no. 1, pp. 485-497, Jan. 2023.

\bibitem{19a}
H. Jing, W. Cheng, W. Zhang, et al., ``OAM based wireless communications with non-coaxial UCA transceiver," \emph{2019 IEEE 30th Annual International Symposium on Personal, Indoor and Mobile Radio Communications (PIMRC)}, pp. 1-6, 2019.

\bibitem{19b}
G. Gil, J. Lee, H. Kim, et al., ``Comparison of UCA-OAM and UCA-MIMO systems for sub-THz band line-of-sight spatial multiplexing transmission," \emph{Journal of Communications and NetworksView article}, vol. 23, no. 2, pp. 83-90, Apr. 2021.

\bibitem{19c}
W. Yu, B. Zhou, Z. Bu, et al., ``UCA based OAM beam steering with high mode isolation," \emph{IEEE Wireless Communications Letters}, vol. 11, no. 5, pp. 977-981, May 2022.

\bibitem{19d}
Z. Guo and G. Yang, ``Radial uniform circular antenna array for dual-mode OAM communication," \emph{IEEE Antennas and Wireless Propagation Letters}, vol. 16, pp. 404-407, 2017.

\bibitem{19e}
F. Qin, L. Li, Y. Liu, et al., ``A four-mode OAM antenna array with equal divergence angle," \emph{IEEE Antennas and Wireless Propagation Letters}, vol. 18, no. 9, pp. 1941-1945, Sept. 2019.

\bibitem{20}
W. Yu, B. Zhou, Z. Bu, et al., ``Analyze UCA based OAM communication from spatial correlation," \emph{IEEE Access}, vol. 8, pp. 194590-194600, 2020.

\bibitem{21}
Y. Yan, G. Xie, M. Lavery, et al., ``High-capacity millimetre-wave communications with orbital angular momentum multiplexing," \emph{Nature Communications}, vol. 5, no. 1, 2014.
\bibitem{22}
Y. Yuan, Y. Zhao, B. Zong, et al., ``Potential key technologies for 6G mobile communications," \emph{Science China Information Sciences}, vol. 5, no. 1, pp. 213-231, 2020.
\bibitem{22a}
W. Long, R. Chen, M. Moretti, et al., ``Joint spatial division and coaxial multiplexing for downlink multi-user OAM wireless backhaul," \emph{IEEE Transactions on Broadcasting}, vol. 67, no. 4, pp. 879-893, Dec. 2021.
\bibitem{22b}
T. Zhang, H. Wang, X. Chu, et al., ``A signaling-based incentive mechanism for device-to-device content sharing in cellular networks," \emph{IEEE Communications Letters}, vol. 21, no. 6, pp. 1377-1380, June 2017.
\bibitem{22c}
Y. Chen, S. He, F. Hou, et al., ``An efficient incentive mechanism for device-to-device multicast communication in cellular networks," \emph{IEEE Transactions on Wireless Communications}, vol. 17, no. 12, pp. 7922-7935, Dec. 2018.
\bibitem{23}
W. Zhang, S. Zheng, Y. Chen, et al., ``Orbital angular momentum-based communications with partial arc sampling receiving," \emph{IEEE Communications Letters}, vol. 20, no. 7, pp. 1381-1384, Jul. 2016.



\end{thebibliography}
\end{document}